\documentclass[prd,aps,floatfix,superscriptaddress,twocolumn,10pt,English]{revtex4}

\usepackage{amssymb,amsmath}
\usepackage{graphicx}
\usepackage[percent]{overpic}
\usepackage{overpic}
\usepackage{color}

\begin{document}  

\title{Testing Gravity with Pulsar Scintillation Measurements} 

\author{Huan Yang}
\affiliation{Perimeter Institute for Theoretical Physics, Waterloo, ON N2L2Y5, Canada}
\affiliation{Institute for Quantum Computing, University of Waterloo, Waterloo, ON N2L3G1, Canada}
\author{Atsushi Nishizawa}
\affiliation{Department of Physics and Astronomy, The University of Mississippi, University, MS 38677, USA}
\affiliation{Theoretical Astrophysics 350-17, California Institute of Technology, Pasadena, California 91125, USA}
\author{Ue-Li Pen}
\affiliation{Perimeter Institute for Theoretical Physics, Waterloo, ON N2L2Y5, Canada}
\affiliation{Canadian Institute for Theoretical Astrophysics, 60 St. George Street, Toronto, On M5S3H8, Canada}
\affiliation{Canadian Institute for Advanced Research, CIFAR program in
Gravitation and Cosmology}
\affiliation{Dunlap Institute for Astronomy \& Astrophysics, University of Toronto, AB 120-50 St. George Street, Toronto, ON M5S 3H4, Canada}

\date{\today}

\begin{abstract} 

We propose to use pulsar scintillation measurements to test predictions
of alternative theories of gravity. Comparing to single-path pulsar
timing measurements, the scintillation measurements can achieve a
part in a thousand accuracy within one wave period, which means pico-second scale resolution in time, due to
the effect of multi-path interference.  
Previous scintillation measurements of  PSR ${\rm B}0834+06$ have data acquisition for hours, making this approach sensitive to mHz gravitational waves. 
Therefore it has unique advantages in measuring gravitational effect or other
mechanisms on light propagation. 
We illustrate its application in constraining scalar gravitational-wave background, in which case the sensitivities can be greatly improved with respect to previous limits. 
We expect much broader applications in testing gravity with existing and future pulsar scintillation observations.

\end{abstract}

\maketitle 

\section{Introduction}
 Pulsar scintillation happens when pulsed radio signals from pulsars follow different paths of propagation to reach the Earth, and exists for almost all known pulsars. It is generally known that structures in interstellar plasma along
the propagation path plays the role of an effective ``lens" and generates
necessary lensing for pulses along different paths to meet at the Earth. Upon arrivals, these radio signals interfere with each other and generate a spatially and frequency varying interference pattern.  
As the Earth is moving, an telescope observer experiences
time-dependent intensity variation corresponding to different fringes
in the interference pattern.  The nature of these lenses is not fully
understood, but appears to be dominated by rare, isolated coherent
plasma structures.  Quantitative models have been proposed to provide
precision templates using a small number of optical caustic parameters\cite{2014MNRAS.442.3338P,2016MNRAS.458.1289L}.

\begin{figure}[t]
  \begin{overpic}[width=0.9\columnwidth,bb=0 0 792 612]{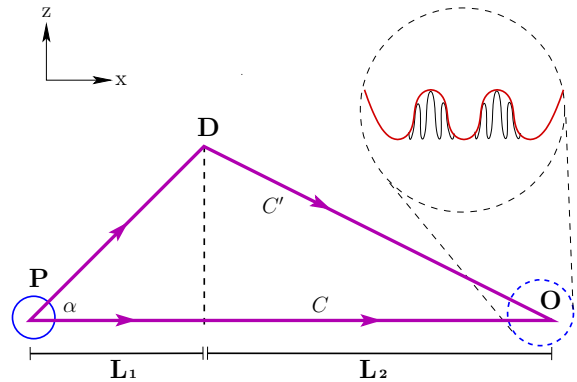}
\end{overpic}
  \caption{(Color Online). The illustration for pulsed signals that
  arrive on the Earth following two distinctive paths, where the wave
  following $\mathcal{C}'$ is deflected by the interstellar medium
  at location ``D". Here $L_1 = r/(1+r) L$ and $L_2 = L/(1+r)$. When the
  radio waves from these two directions reach the observer on the Earth, they
  interfere and produce very fine interference pattern based on the
  radio wavelength $\lambda_e$ and the path opening angle $\alpha
  $. As the Earth moves at a speed $V_e \sim 30\,{\rm km/s}$, there are many
  fringes within the timescale of a single pulse (for illustration
  purpose we only show a few fringes within each pulse).}
\label{fig:pulsarpath}
\end{figure}

As the illustration in Fig.~\ref{fig:pulsarpath}, the spatial
separation between fringes is approximately $\lambda_e/\alpha$
($\lambda_e$ is the radio wavelength, $\alpha$  is the path opening angle) and the temporal separation is $\sim \lambda_e/(\alpha V_e)$, where $V_e$ is the projected
Earth-lens-pulsar velocity, generally dominated by the pulsar proper velocity. 
With $\alpha$ assumed to be $\sim {\rm arcsec}$, one typically observes a scintillation time scale
of seconds, typically longer than the pulsar period. By statistically
(see the discussion in the next section) averaging over time shift of
the fringes, it is possible to achieve phase accuracy that is equivalent to pico-second resolution in time. This is a factor of $10^5$ higher than the accuracy in single-path
pulsar timing \cite{2014MNRAS.440L..36P}. It is worth to note, however, scintillation measurement is fundamentally different from traditional pulsar timing measurements, where the relevant physical quantity in the formal scenario is the radio wave phase differences, and in the latter case it is the pulse arriving time. Therefore it is important to bear in mind that the ``timing precision" in this paper actually refers to the phase accuracies in phase.

This unprecedented phase accuracy (and equivalent timing precision) allows one to apply the scintillation to probing the physics of plasma structures in an interstellar medium \cite{Rickett1977, Rickett1986} and constraining
the size of emission regions in the pulsar magnetospheres
\cite{Johnson2012}. Although high-precision pulsar timing has been
discussed extensively in literature to test alternative theories
of gravity, little was known in relating scintillation measurements
to testing gravity.  In this paper, we propose to use pulsar scintillation
measurements as a laboratory for gravitational physics, in particular,
as a detector of scalar gravitational waves (GWs), which appear in alternative theory of gravity. Similar
analysis can be applied to test other physical  effects that affect
the propagation of radio waves.

\vspace{0.2cm}

\subsection{ Scintillation Modulation}  
 Propagating gravitational distortions modulate the plasma lensing effects. The plasma lenses can change shape on a sound crossing time,
which is typically four orders of magnitude longer than the
gravitational time scales. This allows precise measurements of
space-time variations that are unlikely to be mimicked by plasma
effects. If there exists the GW large enough to be detected, it would lead to an irreducible scintillation model residual.

In the absence of GWs, the variation of the plasma propagation Green's
function is dominated by the Earth-lens-pulsar relative motion.
Interstellar holography retrieves the time dependent Green's
functions, and has been demonstrated to reproduce observed
scintillation patterns to parts per million
\cite{2008MNRAS.388.1214W}.  These authors are able to decompose the
dynamic spectrum as a sum of Green's functions kernel lying
approximately on a parabolic set of loci.  These lenses are located at
a distance of 389pc from earth, with a pulsar distance of
640pc\cite{2016MNRAS.458.1289L}.  The parabolic relationship arises
from the collinearity of lensing points: the time delay through is
lens is proportionate to the square of its transverse separation
angle.  The doppler frequency is the time derivative of this delay due
to the pulsar's apparent motion relative to the screen, and is linear
in transverse separation, thus resulting in a parabolic relationship
of delay and doppler rate of the lensing images.  As a result, plasma lensing induces a modulation frequency proportional to image separation, whereas the change induced by GWs is independent of the separation.
Such a pattern is not observed.  We interpret that achieved
dynamic range of 63dB that no modulation of more than a part in a
thousand in the dynamic spectrum can be due to gravitational waves
moving at the speed of light.  The observing frequency was
approximately 300 MHz, corresponding to wave period $\sim 3 {\rm ns}$ and a Nyquist voltage sampling
rate of $1.5 {\rm ns}$.  We thus estimate the maximum contribution of
gravitational waves at most a part per thousand, or about a picosecond
as the limit on the allowed inverse delay-doppler power. The lower bound of measurable frequency is constraint by the total observation time $t_{\rm obs}$ (for the work in \cite{2008MNRAS.388.1214W}, $1/t_{\rm obs} \sim {\rm m Hz}$). The upper bound of frequency is related to the separation between pulses $1/t_{\rm sep}$, as the pulse sequence determines a natural sampling frequency. A more
precise analysis would require access to the data and holography
algorithm.

The accuracy of this model is limited only by thermal noise, and not
by pulsar self-noise.  A typical $\Delta t \sim$ hour long observation
with $\Delta \nu \sim$ 100 MHz
bandwidth leads to a flux uncertainty of SEFD/$\sqrt{\Delta t\, \delta
  \nu}$, where SEFD is the system equivalent flux density. For large telescopes such as FAST or Arecibo, SEFD is about 5
Jy. 
 There are further subtleties which could affect the sensitivity of scintillation measurement for gravitational waves. First, the $63{\rm dB}$ in power or factor of $10^3$ in signal-to noise-ratio (SNR) is achieved mainly near the bottom of the parabola in the decay time-doppler shift curve, where the signal is the strongest and the effect of GW vanishes. At larger opening angle the data could encode the information GWs but the SNR is lower. Therefore for each specific data set one should try to find the optimum opening delay that balances these two effects. Secondly, it is possible that following the treatment in \cite{2008MNRAS.388.1214W}, part of the noise is absorbed in the model. Therefore it is unclear what fraction of the GW power remains in the residuals. Such fraction may also vary depending on the types of GWs: i.e., single source, continuous/burst sources, GW background, etc.

\vspace{0.2cm}

\section{Probing nontensorial components of GWs} 
 According to the theory of General Relativity, GWs have only two tensor polarizations that are transverse to the wave propagation
direction. 
However, in general metric theory of gravitation \cite{Eardley:1973}, since the metric perturbation $h_{\mu\nu}$ has $10$ components, $4$ of which are purely gauge and eliminated by imposing the condition $h_{0\mu}=0$, there are $6$ degrees of freedom left in $h_{ij}$ ($1 \le i,j \le 3$). 
Therefore gravitational wave emissions with scalar and vector polarizations are predicted in many alternative theories of gravity, such as scalar-tensor theory, $f(R)$ theory, bimetric theory, etc. (For the summary about GW polarization prediction in various alternative gravity models, see \cite{Nishizawa2009} and reference therein). 
Measuring and/or constraining GWs with nontensorial polarizations are a viable approach to test the theories of gravity and search for possible new physics.

We follow the convention in \cite{Nishizawa2009, Lee:2013} to label these $6$ polarizations ($2$ tensor modes: $+$ and $\times$, $2$ vector modes: $x$ and $y$, and $2$ scalar modes: $b$, $l$). In the case that GW is propagating along $z$-axis, the tensor bases  are
\begin{align}
&\tilde{e}^+
= 
\left (
\begin{array}{c c c}
1 & 0 & 0 \\
0 & -1 & 0 \\
0 & 0 & 0
\end{array}
\right ),\quad
\tilde{e}^\times
=\left (
\begin{array}{c c c}
0 & 1 & 0 \\
1 & 0 & 0 \\
0 & 0 & 0
\end{array}
\right )\,, \nonumber \\
&\tilde{e}^b
= 
\left (
\begin{array}{c c c}
1 & 0 & 0 \\
0 & 1 & 0 \\
0 & 0 & 0
\end{array}
\right ),\quad
\tilde{e}^l
=\left (
\begin{array}{c c c}
0 & 0 & 0 \\
0 & 0 & 0 \\
0 & 0 & 1
\end{array}
\right )\,, \nonumber \\
&\tilde{e}^x
= 
\left (
\begin{array}{c c c}
0 & 0 & 1 \\
0 & 0 & 0 \\
1 & 0 & 0
\end{array}
\right ),\quad
\tilde{e}^y
=\left (
\begin{array}{c c c}
0 & 0 & 0 \\
0 & 0 & 1 \\
0 & 1 & 0
\end{array}
\right )\,, 
\end{align}
so that $h_{ij}$ can be decomposed as 
\begin{align}
h_{ij} = \sum_{A} h_A \tilde{e}^A_{ij}\,.
\end{align}  

As an illustration for applications of pulsar scintillation observations to testing gravity, we show that the existing data provide the best constraint on scalar GWB at mHz band, which beats the previous constraint by four orders of magnitude and might be improved by future space-based GW missions such as eLISA \cite{AmaroSeoane:2012km}. 

As shown in Fig. \ref{fig:pulsarpath}, we consider a train of radio
waves emitted from Pulsar (``P") propagates along two different paths
($\mathcal{C}$ and $\mathcal{C}'$) and eventually reaches the Earth.
For simplicity, we consider only one-time deflection by the
turbulent plasma at location "D"(which is straightforward to generalize to cases with multiple deflections), and assume both paths are on the $x-z$ plane, with $\mathcal{C}$ being along $x$ axis. 
The coordinate of ``P", ``D",and ``O" on the $x-z$ plane is $[0,0],[L
r/(1+r), L r \alpha/(1+r)], [L, 0]$ respectively, where $r \equiv L_1/L_2$ in Fig.~\ref{fig:pulsarpath}.

In order to obtain the sensitivity curve to GWs, we derive the transfer functions from GWs with frequency $\omega_g$ in such a system. Based on the standard pulsar timing analysis, the GW-induced phase shift of radio waves propagating along
$\mathcal{C}$  is (hereafter we adopt the geometric unit that the
speed of light $c=1$)
\begin{align}\label{eqc}
H_C = \frac{\pi n^i h_{ij} n^j}{\omega_g \lambda_e} \frac{\sin [\omega_g L \xi+\psi]-\sin \psi}{\xi}\,, 
\end{align}
where $\psi$ is the initial phase of that particular GW, $\xi \equiv
1-{\bf k}\cdot {\bf n}$, with ${\bf k}$ being the unit direction vector
of the GW and ${\bf n}={\bf e}_x$ being the unit direction vector of
$P\rightarrow O$. 

Following the same principle, the phase shift (due to the same GW train)
of radio waves propagating along $\mathcal{C}'$ is
\begin{align}\label{eqcp} H_{\mathcal{C}'}= &\frac{\pi n_1^i h_{ij}
n_1^j}{\omega_g \lambda_e} \frac{\sin [\omega_g r L \xi_1+\psi]-\sin
\psi}{\xi_1}\nonumber\\ &+\frac{\pi n_2^i h_{ij} n_2^j}{\omega_g
\lambda_e} \frac{\sin [\omega_g L \xi+\psi]-\sin [\omega_g r L
\xi_1+\psi]}{\xi_2}\,, \end{align} 
where ${\bf n}_1={\bf e}_x+\alpha \,
{\bf e}_z$, ${\bf n}_2 ={\bf e}_x-r \alpha \, {\bf e}_z$ and
$\xi_{1,2}=1-{\bf n}_{1,2} \cdot {\bf k}$. With $H_{\mathcal{C}}$
and $H_{\mathcal{C}'}$, we can derive the phase shift after averaging over sky directions of the GWs and their initial phases. For example, considering the longitudinal mode, we have
\begin{widetext}
\begin{align}
H'_{\mathcal{C}}-H_{\mathcal{C}} = &\frac{\pi h_l}{\omega_g \lambda_e} \left \{( \sin [\omega_g L_1(1-{\bf n_1}\cdot {\bf k})+\psi]-\sin \psi )\left (\frac{1}{1-{\bf n_1}\cdot {\bf k}}-\frac{1}{1-{\bf n_2}\cdot {\bf k}}\right ) \right .\nonumber \\
& \left .+  (\sin [\omega_g L(1-{\bf k}\cdot {\bf n}+\psi]-\sin \psi)\left (\frac{1}{1-{\bf n_2}\cdot {\bf k}}-\frac{1}{1-{\bf k}\cdot {\bf n}} \right )\right \}\,.
\end{align}

The expression of $(H'_{\mathcal{C}}-H_{\mathcal{C}})^2$ follows obviously. After averaging over the random initial phase $\psi$, and then perform an average over the azimuthal angle around $n$ direction, we arrive at
\begin{align}
(H'_{\mathcal{C}}-H_{\mathcal{C}})^2\approx \frac{\pi^2 h^2_l \alpha^2 (2-\xi)}{2\omega_g^2 \lambda_e^2\xi^3} \left \{  (1-\cos  [\omega L_1 \xi]) \frac{1}{1+r}-(1-\cos  [\omega L\xi]) \frac{r}{(1+r)^2}+(1-\cos  [\omega L_2\xi]) \frac{r}{1+r}\right \}\,.
\end{align}

\end{widetext}
At last the above expression is averaged for $\xi$ (from $0$ to $2$) and that gives the corresponding $\delta \Phi^2$ or
\begin{align}
\delta \Phi =\frac{ \pi h_l \alpha L}{\lambda_e}\sqrt{\frac{r}{2(1+r)}} \sqrt{ \log (1+r)+r\log \frac{1+r}{r}}.
\end{align}

In fact, for other polarizations, we can follow similar procedure to compute the transfer functions.  Their scalings are like
\begin{align}
\delta \Phi^2 &\equiv \langle (H_{\mathcal{C}}-H_{\mathcal{C}'})^2\rangle \nonumber \\
&\propto  \frac{h^2_m \alpha^2}{\omega_g^2 \lambda_e^2} \times \left\{
\begin{array}{cl}
 \log(\omega_g L)\,, & A=+, \times\, b, \\
\\
 \omega_g L\,, & A =x, y\,, \\
\\
\omega_g^2L^2\,, & A=l\,,
\end{array}\right.
\label{bx}
\end{align}
assuming $\omega_g L \gg 1$ (at mHz band it is greater than $10^{8}$
for typical pulsars). In particular, we find that the longitudinal
mode (``$l$") receives the largest amplification factor ($\propto
\omega_g^2 L^2$), while the amplitudes of all other polarizations are suppressed due to the transverse nature of GW propagation. From this reason, here we focus on the longitudinal mode. 

Combining the longitudinal transfer function  with the timing noise estimate given in the previous section, we can obtain the sensitivity of pulsar scintillation measurement on longitudinal scalar GWs, by making $\delta \Phi =2 \pi c\, \delta t_{\rm f}/\lambda_e$. 
Take $r \sim 1$, this gives the sensitivity on $h_l$ (dimensionless GW amplitude) as
\begin{align}
h_l = 6.8\times 10^{-18}  \frac{\delta t_{\rm mHz}}{1\,{\rm ps}}  \left (\frac{\alpha}{{\rm arcsec}} \right )^{-1} \left ( \frac{L}{\rm kpc}\right)^{-1}\,.
\label{eq:scalar-amp}
\end{align}

\begin{figure}[t]
  \begin{overpic}[width=8cm]{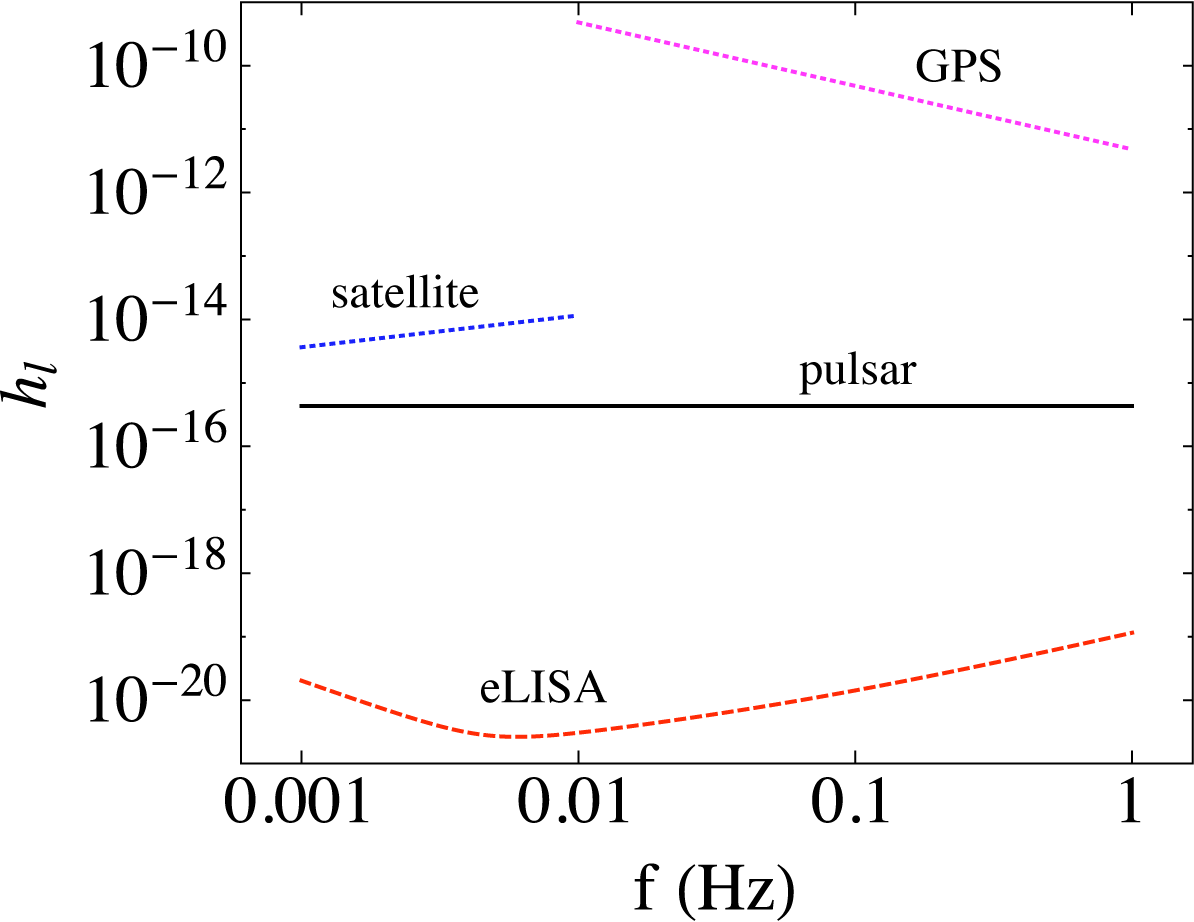}
\end{overpic}
  \caption{(Color Online). The constraint on dimensionless amplitude
  of longitudinal scalar GWs. The lines correspond to sensitivity curves given by previous
  Doppler tracking of spacecraft \cite{Bertotti1995,Armstrong2003} (blue dotted), GPS satellites \cite{Aoyama2014} (magenta dotted), pulsar scintillation from PSR ${\rm B}0834+06$ ($r \sim 0.64, \alpha \sim 20 {\rm mas} /r \sim 31 {\rm mas}$ , and $L \sim 640\,{\rm pc}$ \cite{2016MNRAS.458.1289L,brisken2009100}, black solid), and future eLISA measurements (red
  dashed), respectively. Notice that with the triangular geometry of eLISA, it is difficult to separate out different polarizations of GWs.}
\label{fig:sensitivity}
\end{figure}

In Fig.~\ref{fig:sensitivity}, we compare the sensitivity to
longitudinal scalar GWs based on scintillation measurements of PSR ${\rm B}0834+06$ (from \cite{2008MNRAS.388.1214W, brisken2009100}) with the current best constraint from Doppler tracking of the {\it Cassini} spacecraft in \cite{Bertotti1995,Armstrong2003} and timing measurement of the GPS system in \cite{Aoyama2014} and the proposed sensitivity of eLISA at the same frequency band. As discussed earlier, the SNR of scintillation measurement varies for different opening angle, and in practise the optimal opening angle could be different from the $25 {\rm mas}$ limit obtained in \cite{brisken2009100}. These sensitivities are computed by considering the transfer functions of the scalar longitudinal mode, which give approximately the same responses as the tensor mode for Doppler timings and eLISA below $0.1$ Hz \cite{Tinto2010} but better sensitivity of eLISA above $0.1$ Hz. We can see that scintillation measurement from PSR ${\rm B}0834+06$ already improves the previous sensitivity by a factor of $10$ - $10^6$ (greater improvement comparing to the GPS limit). By choosing more distant pulsars, larger opening angles, and/or the ones with better scintillation timing accuracy, as well as statistically averaging data for different scintillating pulsars,  it is possible to dramatically improve this limit.
\vspace{0.2cm}

\section{Constraint on scalar-tensor ratio of GWs}
 It would be convenient to define the ratio of GW amplitude in scalar mode to that in tensor mode as $R_{\rm ST} \equiv h^{S}/h^{T}$ and useful to show the upper limit in terms of $R_{\rm ST}$. The advantage to use $R_{\rm ST}$ is that it can be interpreted as the relative strength of scalar coupling in a gravity theory to that of the ordinary gravitational (tensor) coupling, because the ratio is irrespective of common factors between the scalar and tensor modes, e.g. distance to the source and the way of propagation in the interstellar space. It should be emphasized that in general in modified gravity theory, the scalar coupling strength depends on an environment in the Universe, so called the screening mechanism, e.g. the Chameleon mechanism, the Vainshtein mechanism, and etc. \cite{Khoury:2004, Vainshtein1972}. Our constraint is obtained in a low-density and weak-gravity region (in cosmological sense). In a high-density and stronger-gravity region such as near a GW source or on the Earth,  relatively large deviation from general relativity is allowed where screening mechanism is also likely to operate. However, that part of contribution is highly model-dependent.

To derive the upper limit on $R_{\rm ST}$, what we need is the upper limit on the scalar amplitude in Eq.~(\ref{eq:scalar-amp}) and the amplitude in the tensor mode. The latter is source-dependent and has large uncertainty, depending on astrophysical scenarios. Thus we take into account this uncertainty, adopting the lowest, intermediate, and highest event rates among predictions in literature when we derive the power spectrum densities $S_h$ of each GW source. 
For white dwarf (WD)  binaries, the extragalactic component dominates at $f > 1\,{\rm mHz}$ and the spectrum has been estimated in \cite{Farmer2003MNRAS} as $S_h^{\rm{WD}}(f) = \{0.37,1.4,2.3\} \times 10^{-46} \left( f/{\rm Hz} \right)^{-7/3} \exp \left[- f/0.01\,{\rm Hz} \right] \;\; {\rm{Hz}}^{-1}$, each corresponding to the lowest, intermediate, and highest event rates. 
For neutron star (NS) binaries, compiling the present merger rate \cite{Abadie:2010cf} and its redshift evolution \cite{Cutler:2006} gives $S_h^{\rm{NS}} (f) = \{0.016, 1.6, 16 \} \times 10^{-47}\left( f/1\,{\rm Hz} \right)^{-7/3}$ below a kHz band. 
For black hole (BH) binaries, the recent detection of a massive BH binary indicates that the merger rate of BH binaries may be higher than the previous expectations \cite{GW150914:GWB}. Although the power spectrum of the GWB depends on models of BH binary formation, the {\it fiducial} model in \cite{GW150914:GWB} gives $S_h^{\rm{BH}} (f) = \{0.86,4.7,16\} \times 10^{-47}\left( f/1\,{\rm Hz} \right)^{-7/3}$ without a high frequency cutoff in our interest frequency band. 

\begin{figure}[t]
\begin{center}
\includegraphics[width=8.2cm]{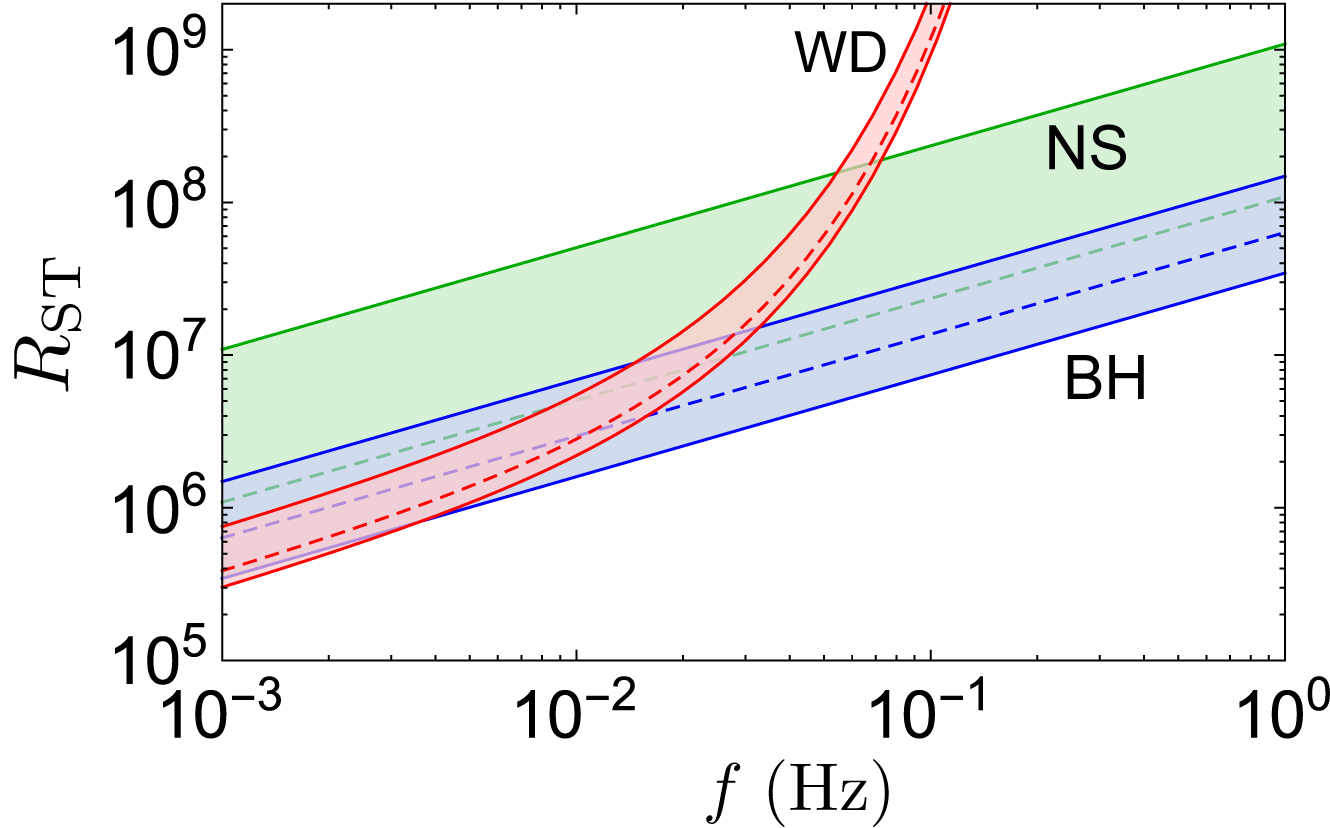}
\caption{(Color Online). Upper limit on the scalar-to-tensor ratio $R_{\rm ST}$ for each GWB source with uncertainties of merger rates: WD binaries (red), NS binaries (green), and BH binaries (blue). The dashed lines correspond to the intermediate merger rates and the solid lines are the lowest and highest merger rates.}
\label{fig:ST-ratio}
\end{center}
\end{figure}

In Fig.~\ref{fig:ST-ratio}, the constraints on $R_{\rm ST}$ for each GW source are shown. The upper bound is tighter at lower frequencies, which are at $1\,{\rm mHz}$, $3.9 \times 10^5$, $1.1 \times 10^6$, and $6.4 \times 10^5$ for the intermediate merger rates of WD, NS, and BH binaries, respectively. Although the numerical values appear to be much larger than one, they are the first constraints on $R_{\rm ST}$ obtained in the frequency band from $1\,{\rm mHz}$ to $1\,{\rm Hz}$ in the low-density and weak-gravity region of space, and they connect the physics of GW emission of a source and a screening mechanism in a model-independent way. There have been the constraints at different frequencies from other observations. The observation of the orbital-period derivative from PSR B1913+16 agrees well with predicted values of GR, conservatively, at a level of $1\,\%$ error \cite{Weisberg2010ApJ}. This fact indicates that the contribution of scalar GWs to the energy loss is less than $1\,\%$, that is, $R_{\rm{ST}} \lesssim 10^{-1}$ at $7.2 \times 10^{-5}\,{\rm Hz}$ at the source position of the NS binary. On the other hand, the recent detection of GWs (GW150914) \cite{GW150914PRL} gives no constraint on the scalar component, as at least three detectors are needed to break the degeneracy of the polarization modes \cite{Nishizawa:2009jh}.

\section{Discussion and Conclusion}
 Comparing to single path pulsar timing
measurements, the scintillation measurements have better timing
accuracies, and the phase-comparison geometry which naturally removes
intrinsic noise from the source. These are the key factors which
ensures its ultra precision and enables its application to studying
ISM physics, pulsar physics, and our proposal in this paper - testing
alternative gravity models.

We have illustrated an example in this proposal: measuring a
longitudinal scalar GWB. It is also possible
to apply to other tests which do not involve GWs - for example, the spacetime quantum fluctuations \cite{Ng1994MPLA,Amelino-Camelia1999Nature} or
the holographic noise \cite{Hogan2008PRD}. They would contribute distinctive phase noise for photon traveling along different scintillation paths, and hence can be measured by observing
anomalous scintillation phase shift or degrading of the interference
pattern.

\acknowledgements
The authors appreciate many helpful comments from the referees, especially regarding many aspects of the scintillation discussions. HY thank I-Sheng Yang for very instructive
discussions on timing noise of Pulsar scintillations and Nestor
Ortiz for making Fig.1. HY acknowledges supports from the Perimeter
Institute of Theoretical Physics and the Institute for Quantum
Computing. AN are supported by NSF CAREER Grant No. PHY-1055103 and the H2020-MSCA-RISE- 2015 Grant No. StronGrHEP-690904. AN thanks the hospitality of Perimeter Institute, where part of the work was performed. Research at Perimeter Institute is supported by the government of Canada  through the Department of Innovation, Science and Economic Development Canada and by the Province of Ontario though Ministry of Research and Innovation.

\bibliography{References}

\end{document}